\documentclass[prl,twocolumn,draft,amsmath,showpacs]{revtex4}
\usepackage{graphics}\input epsf
\begin{document}

\title{Charge density waves in Sr$_{14-x}$Ca$_{x}$Cu$_{24}$O$_{41}$: electron
correlations vs. structural effects}
\author{
 K.-Y. Choi,$^{1,2}$
 M. Grove,$^{1,3}$
 P. Lemmens,$^{4,5\dag}$
 M. Fischer,$^1$
 G. G\"untherodt,$^1$
 U. Ammerahl,$^6$
 B. B\"uchner,$^7$
 G. Dhalenne,$^8$
 A. Revcolevschi,$^8$
 and J. Akimitsu$^9$}

\affiliation{$^1$ 2. Physikalisches Institut, RWTH-Aachen, 52056
Aachen, Germany}
 \affiliation{$^2$ Institute for Materials Research, Tohoku
University, Katahira 2-1-1, Sendai 980-8577, Japan}
 \affiliation{$^3$ Philips Automotive Lighting,
 D-52068 Aachen, Germany}
 \affiliation{$^4$ Institute for Physics of
Condensed Matter, TU Braunschweig, D-38106 Braunschweig, Germany}
 \affiliation{$^5$ Max Planck Institute for Solid State Research,
 D-70569 Stuttgart, Germany}
 \affiliation{$^6$ 2. Physikalisches Institut, Universit\"at zu
K\"oln, D-50937 K\"oln, Germany }
 \affiliation{$^7$ Institute for
Solid State Research, IFW Dresden, D-01171 Dresden, Germany}
 \affiliation{$^8$ Laboratoire de Physico-Chimie, Universit\'{e}
Paris-Sud, B\^at. 414, 91405 Orsay, France}
 \affiliation{$^9$
Department of Physics, Aoyama Gakuin University, Tokyo 157-8572,
Japan}

\date{\today}

\pacs{78.30.–j, 71.27.+a, 71.45.Lr}

\begin{abstract}
We report inelastic light scattering experiments in the spin-chain and ladder system
Sr$_{14-x}$Ca$_{x}$Cu$_{24}$O$_{41}$. A depletion of an electronic background is
observed for $0\leq x \leq5$ at low temperatures and with light polarization both
parallel to the leg and rung directions. This points to the formation of 2D charge
density wave in the ladders. Furthermore, the Ca-substitution dependence of the 305- and
249-cm$^{-1}$ modes, sensitive to the lattice modulations of the respective chain and
ladder system, suggests that the charge dynamics in the ladders is strongly influenced
by the structural incommensurability.
\end{abstract}

%}

\maketitle

\narrowtext

%\newpage

The spin-chain and ladder compound
Sr$_{14-x}$Ca$_{x}$Cu$_{24}$O$_{41}$ (SCCO) has attracted intense
theoretical and experimental attention over the last years
\cite{Dagotto99}. This interest is triggered by the observation of
a superconducting phase for $x=13.6$ under pressure
\cite{Uehara96} and an unconventional charge density wave (CDW)
state for $x<9$
\cite{Blumberg02,Gozar03,Gorshunov02,Vuletic03,Vuletic05,Abbamonte04,Rusydi05}.

SCCO is built up by alternating layers of edge-sharing $\rm CuO_2$
chains and two-leg $\rm Cu_2O_3$ ladders in the $ac$ plane
\cite{Mccarron88}. These two sublayers are separated by (Sr,Ca)
atoms, and stacked along the $b$ axis. The chains and ladders run
along the $c$ axis and their unit cell parameters differ by a
factor of $\sqrt{2}$ along this direction. This misfit leads to an
incommensurate modulation of the two sublattices. In addition,
SCCO is intrinsically ($x=0$) hole-doped with a formal Cu valence
of 2.25.

For $x=0$ the chain system shows a dimerization and charge
ordering for temperatures below 200 K \cite{Cox98,Takigawa98}.
Specific modulations of the chain lattice are responsible for the
formation of the dimers between two copper spins separated by one
Zhang-Rice singlet \cite{Cox98,Takigawa98,Matsuda99}. The
substitution of Sr by isovalent Ca leads to the instability of the
dimerization and charge ordering \cite{Kataev01}. Concerning the
electronic state of the ladder, several spectroscopy studies
\cite{Blumberg02,Gozar03,Gorshunov02,Vuletic03,Vuletic05,Abbamonte04,Rusydi05}
provide evidence for a CDW. In detail, however, there is a
profound disagreement in the interpretation of the experimental
data. Far-infrared (IR) results
\cite{Gorshunov02,Vuletic03,Vuletic05} show that the insulating
CDW is two dimensional with an anisotropic dispersion and vanishes
for $x>9$. In contrast, a dynamic Raman response assigned to CDW
fluctuations is shown to persist in the metallic phase of $x=12$
\cite{Gozar03}. Very recently, resonant X-ray scattering
measurements \cite{Abbamonte04,Rusydi05} reported the observation
of a hole crystal at $x=0, 10, 11,$ and 12 with a commensurate
wave vector, which is attributed to electronic correlations
\cite{Dagotto92,White02}. Noticeably, the hole crystal is absent
with intermediate dopings $1\leq x \leq5$. The conflicting
spectroscopy results highlight the unusual electronic and
structural properties of the ladder and point to an intimate
interplay between holes and lattice modulations as recent
experimental and theoretical works suggest
\cite{Gelle04,Gelle04b,Braden04,Zimmermann04}.

Up to now, several Raman measurements have been reported on magnetic and charge
excitations while focusing mainly on either the high-frequency or the low-frequency
regime \cite{Blumberg02,Gozar03,Popovic99,Sugai99,lemmens00}. The intermediate frequency
regime has not been carefully investigated in spite of its potential relevance to the
charge ordered state in the ladders.

In this study, we report Raman scattering measurements of
Sr$_{14-x}$Ca$_{x}$Cu$_{24}$O$_{41}$. We highlight a pseudogap behavior of an electronic
background for $0\leq x \leq 5$ with incident and scattered light scattering
polarizations both parallel and perpendicular to the ladder direction and its
disappearance for $x=12$. This is attributed to the formation of a 2D CDW. In addition,
we found that the substitution dependence of the CDW is closely associated with that of
charge ordering in the chain system by accompanying lattice modulations of both
subsystems. This suggests that the CDW relies on the mutual interaction of charge
correlations and lattice incommensurability.

Single crystals were grown by a floating zone method~\cite{Ammerahl99}. The experiments
were performed using the $\lambda = 514.5$~nm excitation line of an Ar$^+$ ion laser
(laser fluency less than 20~W/cm$^2$). Thus, the incident radiation did not increase the
temperature of the samples by more than 1~K. The scattered spectra were collected by a
DILOR-XY triple spectrometer and a back-illuminated CCD detector in a
quasibackscattering geometry.

\begin{figure}[th]
      \begin{center}
       \leavevmode
       \epsfxsize=8cm \epsfbox{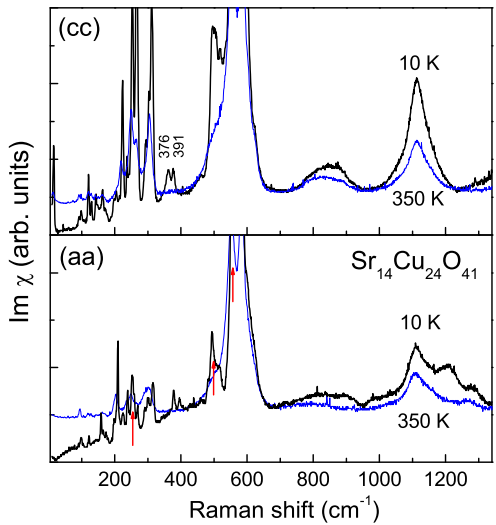}
        \caption{(online color) Raman spectra of $\rm Sr_{14}Cu_{24}O_{41}$ in
($cc$) and ($aa$) polarizations at 10~K and 350~K, respectively. A
strong renormalization is observed in the electronic background
for frequencies below 400 cm$^{-1}$ at low temperatures.}
\label{SCOf1}
\end{center}
\end{figure}

Raman spectra are displayed in Fig.~1 in ($cc$) and ($aa$) polarizations at 10 K and 350
K each. At 350 K we can resolve 31 peaks as one-phonon scattering in the frequency range
up to 700~cm$^{-1}$. The broad peaks observed between 700-1300~cm$^{-1}$ are due to
two-phonon scattering. A factor group analysis based on the independent chain (Cmcm) and
the ladder (Fmmm) subsystem yields
$\Gamma=6A_g(aa,bb,cc)+6B_{1g}(ab)+4B_{2g}(ac)+2B_{3g}(bc)$ \cite{Popovic99}.  The extra
13 modes should be attributed to the composite unit cell with the crystal symmetry of
Pcc2. With decreasing temperature all peaks become sharp and grow considerably in
intensity, suggesting an enhanced localization of charges, i.e. holes. Remarkably, the
intense 376- and 391-cm$^{-1}$ modes appear below the charge ordering temperature at 200
K. This implies that they do not rely on the chain and ladder incommensurability.
Furthermore, they are not zone-folded modes because they have no counterpart modes at
high temperatures and have strong intensity in contrast to an usual zone-folded mode.
Therefore, we propose that they are related to a structural symmetry change around 200 K
as a neutron diffraction study indicates \cite{Braden04}. Our result suggests that
charge ordering leads to an enhanced, additional deformation of the lattice. In the
literature \cite{Popovic99,Sugai99,lemmens00} the origin of several peaks at 269, 496,
and 569~cm$^{-1}$ (indicated by arrows) is controversially discussed in terms of
magnetic excitations. However, we assign them to phonons. This is because (i) in an
external field they exhibit neither a shift of the peak energy nor a splitting, and (ii)
the observed narrow peaks are not compatible with the pronounced dispersion of the
triplet excitations \cite{Schmidt01}.

The spectra show an important feature at low temperatures and for frequencies below
about 400 cm$^{-1}$. There is a depletion of a continuous scattering background
attributed to electronic scattering. Such a pseudogap behavior is typical for a charge
ordered state in strongly correlated electron systems. To further substantiate this
effect we will use specific phonon modes to probe the charge dynamics. The high
sensitivity of phonons to charge disproportionation is due to the related modulation of
lattice force constants \cite{Fischer99,Sherman99}. In SCCO, the $249$- and
$305$-cm$^{-1}$ modes correspond to vibrations of the Cu ladder and chain atoms,
respectively \cite{Popovic99}. This enables us to address charge ordering within the two
sublattices separately.

\begin{figure}[th]
      \begin{center}
       \leavevmode
       \epsfxsize=8cm \epsfbox{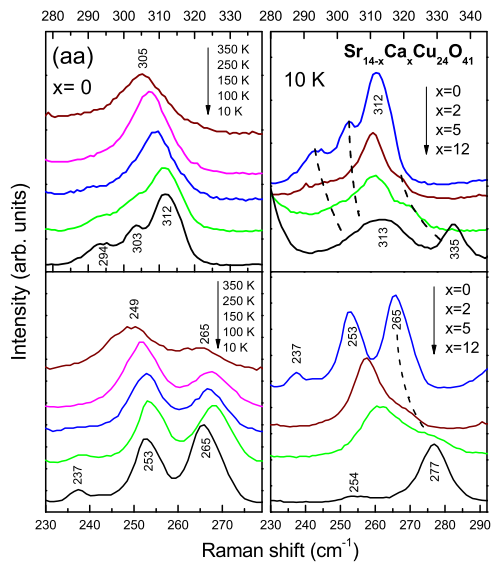}
        \caption{(online color) (left panel) Temperature dependence
        of the $305$- and $249$-cm$^{-1}$ modes which correspond to vibrations
        of the Cu chain and ladder atoms, respectively. (right panel)
        Substitution dependence of the $305$- and $249$-cm$^{-1}$ modes at 10~K.
        All spectra were measured in (aa) polarization. }
\label{SCOf2}
\end{center}
\end{figure}

Figure 2 displays a zoom of the temperature (350 to 10 K) and substitution dependence
($0<x<12$) at 10 K of specific phonon modes in (aa) light scattering polarization. With
decreasing temperature from 350 K to 10 K the $305$-cm$^{-1}$ chain mode hardens by
$7$~cm$^{-1}$ while additional split-off modes show up at $294$~cm$^{-1}$ and
$303$~cm$^{-1}$ [see the left upper panel]. The latter correspond to zone-folding
effects due to the formation of a superstructure. This confirms the presence of charge
ordering in the chain. With increasing Ca-content the split-off modes are suppressed
while the $312$-cm$^{-1}$ mode broadens [see the right upper panel].  A distinct feature
is that at the same time a new mode at the right shoulder of the main peak appears. This
mode is more pronounced and observed as a separate peak $335$~cm$^{-1}$ for $x=12$.
Thus, the gradual switching of the spectra from $x=0$ and $x=12$ can be ascribed to a
continuous change of the lattice modulations leading to a suppression of the charge
ordering. Note that only one broadened peak around 313 cm$^{-1}$ is expected in the case
of a homogeneous charge distribution. In this light, an extra peak at the right shoulder
implies an inhomogeneous charge distribution at high $x$. This is confirmed by ESR
measurements \cite{Kataev01}.

Compared to the chain mode, the $249$-cm$^{-1}$ phonon of the ladder system shows a less
pronounced hardening by $4$~cm$^{-1}$ [see the left lower panel]. Furthermore, we find
no evidence for zone-folded modes. Instead, upon cooling all modes gain a strong
intensity while the $237$-cm$^{-1}$ mode appears below 150~K. This suggests that the
charge ordering in the ladder is weakly coupled to lattice modulations. With increasing
Ca-content the $253$-cm$^{-1}$  at 10 K becomes broader and weaker. [see the right lower
panel]. This is related to the weakening of the CDW in the ladder [see Fig. 4 and
below]. Noticeably, the doping dependence of the ladder mode resembles that of the chain
mode because of an incommensurate lattice modulation of the two sublattices.

%$305$-cm$^{-1}$ (350 K)

Shown in Fig. 3(a) is the temperature dependence of the peak position of the
$305$-cm$^{-1}$ chain mode as a function of Ca-content. To identify the charge ordering
contribution to the frequency shift, we have estimated the anharmonic phonon
contribution using a model based on phonon-phonon decay processes
\cite{choi03,Balkanski83},
\begin{equation}
\omega_{ph}(T)=\omega_0 + C[1+2/(e^{x}-1)],
\end{equation}
where $x=\hbar\omega_0/2k_BT$ with the value of $\omega_0=312$ cm$^{-1}$ and $C=-1.4$
cm$^{-1}$ for $x=12$. The $x=12$ sample shows good agreement between the fitted and
experimental data in the overall temperature range. This means that the temperature
dependence of the phonon frequency is solely due to lattice anharmonicity. With
decreasing $x$ the deviations show up and develop in a systematic way. This gives
evidence for the presence of other contributions in addition to anharmonic lattice
effects. Here we mention that for small $x$ the fitting interval is restricted to the
high temperature range where a linear temperature dependence is observed. Remarkably,
for $x=0$ the temperature dependence of the frequency shift below $T^{*}=200~$K is
similar to that of the X-ray scattering intensity of the superlattice reflections at
(0,0,$l$) positions \cite{Cox98,Zimmermann04}. Thus, we identify the characteristic
temperature, $T^{*}$, where the discrepancy starts to appear as the onset of charge
ordering. With increasing Ca-content $T^{*}$ shifts to lower temperature [see Fig. 3(a)
and (b)].

These observations can be explained as follows: At high temperature the frequency of the
O-vibration within the $\rm CuO_4$ plaquette is given by $\nu_0=\sqrt{\kappa/M}$ with
$\kappa$ the lattice force constant and $M$ the O-mass. At low temperatures charge
ordering leads to an additional modulation of the lattice. Thus, the lattice force
constant changes due to Coulomb repulsion between the Cu sites $\delta_{Cou}$.
Therefore, the renormalized frequency shift is given by
$\nu=\sqrt{(\kappa+\delta_{Cou})/M}$. In this regard, the decrease of the deviation
below $T^{*}$ with increasing $x$ implies that the charge ordering induced lattice
distortion weakens. However, this does not mean that the chain and the ladder become
homogeneous. As discussed above, the structural modulations are still substantial at
$x=12$ [see the right panel of Fig. 2].

\begin{figure}[th]
      \begin{center}
       \leavevmode
       \epsfxsize=8.5cm \epsfbox{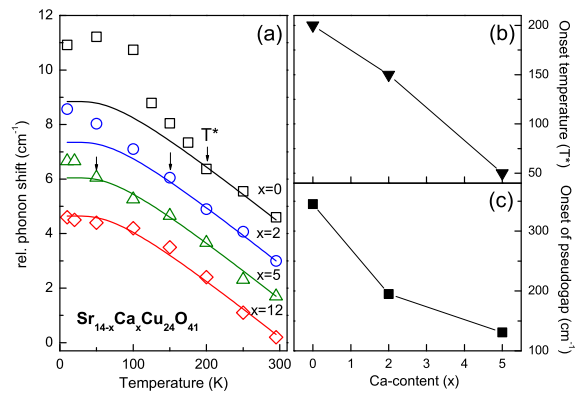}
        \caption{(online color)(a) Temperature and substitution dependence
         of the peak position of the 312-cm$^{-1}$
        mode. The solid lines represent a fitting by eq. (1).
        The characteristic temperature, $T^{*}$, is indicated by an arrow. (b)
        $T^{*}$ as a function of Ca-content. (c) The onset frequency of the
        pseudogap as a function of Ca-content.} \label{SCOf3}
\end{center}
\end{figure}

We will now turn again to the low-temperature depletion of the electronic background
observed for frequencies below 400~cm$^{-1}$ in both ($cc$) and ($aa$) polarizations
[see Fig. 1]. Figure 4 displays its substitution dependence at 10 K in ($cc$)
polarization. Since the background level is constant in the frequency range 400 -1400
cm$^{-1}$, we use the intensity between 400-450 cm$^{-1}$ as a reference point for
$x>0$. The ($aa$) polarization displays a similar behavior (not shown here). With
increasing $x$ the onset energy of the depletion shifts to lower energy (indicated by
the vertical arrow) and the pseudogap closes and finally disappears for $x=12$. The
suppression of spectral weight can be taken as a definite signature of a CDW in the
ladder system. Such a pseudogap behavior is characteristic for a charge ordered state in
strongly correlated electron systems. This points to the significant role of electronic
correlations in forming the CDW state.

Actually, Raman, IR, and x-ray scattering studies
\cite{Blumberg02,Gozar03,Gorshunov02,Vuletic03,Vuletic05} provide evidence for the
presence of such a modulated state. There are, however, differences in details derived
from these experimental techniques. In Raman scattering measurements
\cite{Blumberg02,Gozar03} low energy quasielastic scattering (QES) is attributed to
collective CDW excitations. An analysis of this dynamic Raman response seems to indicate
that a 1D local CDW exists for all hole concentrations. In contrast, recent optical
absorption spectra \cite{Vuletic05} show (i) a CDW response along both the leg and rung,
(ii) the vanishing of the CDW signal along the rung at $x=8$, and (iii) the
disappearance of the CDW state for $x>9$. This would imply a 2D nature of the observed
CDW state. In 1D electron systems electronic Raman response which is proportional to a
curvature of dispersion is possible only in one polarization direction. Therefore, the
observation of the electronic response in both (aa) and (cc) polarization evidences the
2D nature of the CDW state for Sr concentrations at least up to $x=5$. Thus, our Raman
results with respect to the polarization and doping dependence are fully consistent with
these IR results. The 2D CDW is further supported by x-ray scattering measurements which
shows a coherent modulation across $\sim 50$ neighboring ladders \cite{Abbamonte04}.

\begin{figure}[th]
      \begin{center}
       \leavevmode
       \epsfxsize=8.5cm \epsfbox{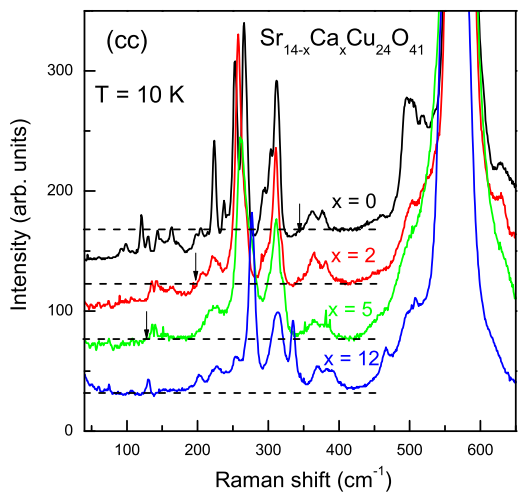}
        \caption{(online color) Low temperature Raman spectra of Ca substituted $\rm Sr_{14-{\it
x}}Ca_{\it x}Cu_{24}O_{41}$ with $0\leq x\leq12$ for intraladder ($cc$) polarization.
The dashed lines correspond to a reference point with respect to the unrenormalized
electronic continuum of each spectrum and the arrows mark the onset of spectral weight
depletion, i.e. opening of the pseudo gap.} \label{SCOf4}
\end{center}
\end{figure}

The conflict in the derived pictures is in part related to the complexity of the present
system with chains, ladders and coupled spin/charge degrees of freedom. The main
problem, however, is based on the ambiguity of interpreting QES in terms of CDW
excitations. Usually, an unspecific hydrodynamic model is adopted to analyze the QES
with respect to electronic scattering from spins, impurities, and phonons. ESR
measurements \cite{Kataev01} evidence the significance of mobile holes in the chain for
all $x$ above the charge ordering temperature. In this case, a scattering of mobile
holes by spins in the chain contributes to a QES. Actually, QES is observed (i) in a
wide concentration range, $0<x<12$, (ii) only for polarization parallel to the leg, and
(iii) in the high temperature regime. Therefore, we come to the conclusion that the
observed QES in earlier Raman studies \cite{Blumberg02,Gozar03}, might be governed by
the charge dynamics of the chain rather than the CDW of the ladder.

In the following we will discuss the origin of the CDW order. Although there is some
evidence for a conventional CDW \cite{Vuletic03}, the presented Raman spectra show no
characteristic features which are expected for a conventional Peierls mechanism. That
is, there is no clear hint for the presence of a Raman-active amplitude mode.
Furthermore, no significant lattice modulations are accompanied by the formation of the
CDW [see the left lower panel of Fig. 2]. However, a Peierls mechanism cannot entirely
be ruled out because the CDW-related superstructure might be negligibly weak and
difficult to observe by Raman spectroscopy. Further high resolution x-ray or electron
diffraction studies might be more decisive. Here, we want to recall that a recent
resonating X-ray scattering experiment \cite{Abbamonte04} uncover the important role of
electronic correlations as the driving force to the CDW. If considering very weak
electron-phonon interactions, the CDW-related pseudogap feature can be ascribed to
electronic correlation effects. Noticeably, the substitution dependence of the CDW
resembles that of the charge ordering in the chain [compare Figs. 3 (b) and (c)]. Thus,
the full aspect of the CDW is described by the mutual interaction between the hole
localization and a lattice modulation.

In a qualitative picture, the hole-lattice interactions can be explained as follows: The
chain-lattice distortions lead to a pinning potential for the holes in the chain. In
turn, the localized holes deform the lattice. The ladder-lattice is modulated according
to the chain-lattice distortions. This influences the localization pattern of the holes
in the ladder. In the same way, charge ordering in the ladder affects the chain-lattice
modulation. We emphasize that recently increasing evidence exists for such an intrinsic
interplay between the charge ordering and the lattice modulations from both experimental
and theoretical sides \cite{Braden04,Zimmermann04,Gelle04,Gelle04b}. The controllability
of the charge ordering by hole-lattice interactions has an analogy in the stripe
scenario of high-T$_c$ single layer cuprates where the cooperative octahedral tilts
induce a pinning potential for charge stripes \cite{Simovic}.

In summary, an intimate connection between the charge dynamics and the incommensurate
lattice modulation in Sr$_{14-x}$Ca$_{x}$Cu$_{24}$O$_{41}$ has been discussed. Evidence
has been presented that the 2D CDW in the ladder system and the charge ordering in the
chain system are based on an intriguing interplay between specific lattice modulations
and electronic correlations. Our study highlights the significance of inhomogeneous
lattice effects in understanding the electronic properties of
Sr$_{14-x}$Ca$_{x}$Cu$_{24}$O$_{41}$.

This work was supported by the German Science Foundation, DFG/SPP 1077 and the European
Science Foundation "Highly Frustrated Magnetism". We acknowledge important discussion
with P.H.M. van Loosdrecht, T. Kopp, V. Kotov, G. Uhrig, and V.
Torgashev.\\

$^\dag$To whom correspondence should be addressed. E-mail: p.lemmens@tu-bs.de.

\end{document}